\documentclass[aps,twocolumn,superscriptaddress]{revtex4-2}
\usepackage{amsmath}
\usepackage{bm}
\usepackage{graphicx}
\usepackage{amssymb}
\usepackage{xcolor}

\begin{document}

\title{Anisotropic Photon Emission Enhancement near Carbon Nanotube Metasurfaces}

\author{Michael D. Pugh}
\affiliation{Department of Mathematics \& Physics, North Carolina Central University, Durham, NC 27707, USA}

\author{SK Firoz Islam}
\affiliation{Department of Mathematics \& Physics, North Carolina Central University, Durham, NC 27707, USA}
\affiliation{Department of Physics, Jamia Millia Islamia (A Central University), New Delhi 110025, India}

\author{Igor V. Bondarev}\email[Corresponding author: ]{ibondarev@nccu.edu}
\affiliation{Department of Mathematics \& Physics, North Carolina Central University, Durham, NC 27707, USA}

\begin{abstract}
We present a theoretical study of the directionality effects in spontaneous emission and resonance fluorescence of a quantum two-level dipole emitter near an ultrathin closely packed periodically aligned single-wall carbon nanotube film. Such films present an example of highly anisotropic flexible metasurfaces that are now available experimentally. The nanotube alignment is shown to provide an extra measure for quantum control of dipolar spontaneous emission and resonance fluorescence in such systems, in addition to film thickness and composition parameters such as tube diameter, chirality and translational period. The processes studied are shown to be highly anisotropic, being enhanced by orders of magnitude in the direction perpendicular to the alignment and metasurface plane, contrasting with the commonly believed viewpoint of their uncontrollably random directionality.
\end{abstract}

\maketitle

\section{Introduction}

A carbon nanotube (CN) is a graphene sheet rolled into a cylinder such that its length far exceeds its diameter of a few nanometers~\cite{Saito1998,Doorn04,Baughman13}. Its helical and translational symmetries are specified with two chiral indices that indicate how the edges of the graphene lattice are bonded at the seam. These symmetries in turn establish the tube's axial electrical conductivity, which can range from metallic to semiconducting. Properties of CNs may be tuned by adjusting their diameter, chirality and, as with other semiconductors, by doping. While this tunability makes CNs highly effective on their own in areas ranging from electron transport~\cite{MArnold21,GeBo2016,Ando2005} to spectroscopy related electromagnetic (EM) response phenomena~\cite{Hage2017,Zaumseil2022,Gao2018,Bondarev2017,Graf2016,Bondarev2015a,MartinMoreno2015,Bondarev2014,Bondarev2014a,Zhang2013,Bond2012,BondAnt2012,PoWoodsBond11,Bondarev2009,Dresselhaus2007,BondLamb06}, it also compels their use in composite transdimensional (TD) material systems such as planar periodic arrays and films~\cite{MArnold22,Brady2016,Liu2020,Roberts2020,Roberts2022,Roberts2020,Falk2020,Roberts2019,Naik2019,Kono2019,Falk2018,Falk2017}. TD quantum materials are ultrathin films made of precisely controlled number of monolayers~\cite{Boltasseva2019,Shah2020}. Whereas three-dimensional (3D) bulk materials allow for higher free carrier concentration and their two-dimensional (2D) counterparts such as graphene and monolayer transition metal dichalcogenides provide the strong confinement of exciton-polariton and plasmon modes~\citep{Basov2014,Mak2016,Xia2014}, the advantages of both of these extremes can be merged by using TD materials~\cite{Shen2023,BiehsBond2023,Bond2022,Shah2022,Manjavacas2022,Bondarev2020,Pruneri2019,Abajo2019,Bondarev2018,Bondarev2017a,Campione2015}. For periodic CN arrays and films, in addition to controlling their thickness, new tunable attributes emerge: tube-diameter-to-film-thickness ratio, intertube spacing, and abundance ratio of tube geometries in the array assortment~\cite{Bondarev2023,Pablo2024,Adhikari2021,Bondarev2021,Bondarev2019}.

In recent years, modern fabrication techniques have appreciably improved the quality of CN films~\cite{Komatsu2020,Rust2022,Zhukov2022}. This has led to a myriad of diverse applications such as single-photon sources~\cite{Zaumseil2022,Zheng2022,Li2022,Saha2018,Khasminskaya2016,Ma2015}, field effect transistors~\cite{Brady2016,Liu2020}, rechargeable batteries~\cite{Fan2018}, electrothermal actuators~\cite{Ghosh2020}, electron~\cite{Su2023} and thermal~\cite{Roberts2022} transport control, super\-capacitors~\cite{Zhang2017}, solar cells~\cite{Qiu2015}---even directional dark matter detection~\cite{Pandolfi2021} and agriculture~\cite{Wang2019}---to name a few. A significant sector of current TD materials research is committed to experiment and theory to study self-assembled thin films of periodically aligned single-wall carbon nanotube (SWCN) arrays~\cite{Roberts2022,Roberts2020,Falk2020,Roberts2019,Naik2019}---a new highly anisotropic TD material platform for designing efficient flexible unidirectional hyperbolic metasurfaces with characteristics adjustable on demand by means of their thickness, SWCN diameter, chirality, and periodicity variation. It was recently shown experimentally~\cite{Roberts2019} and explained theoretically~\cite{Bondarev2021} that in the SWCN alignment direction the real part of the linear EM response function of the ultrathin single-type CN array has a broad negative refraction band near a quantum interband transition of the constituent SWCN, whereby the film behaves as an in-plane anisotropic (unidirectional) hyperbolic metasurface at much higher frequencies than those (typically in the IR~\cite{Chen20}) that classical plasma oscillations have to offer. By decreasing the CN diameter it is possible to push this negative refraction band into the visible range and using weakly inhomogeneous doped multitype SWCN films broadens the negative refraction bandwidth~\cite{Bondarev2021}, to allow both near- and far-field interaction control in the SWCN array systems~\cite{Bondarev2023,Pablo2024,Adhikari2021}.

\begin{figure}[t]
\includegraphics[width=1.0\linewidth]{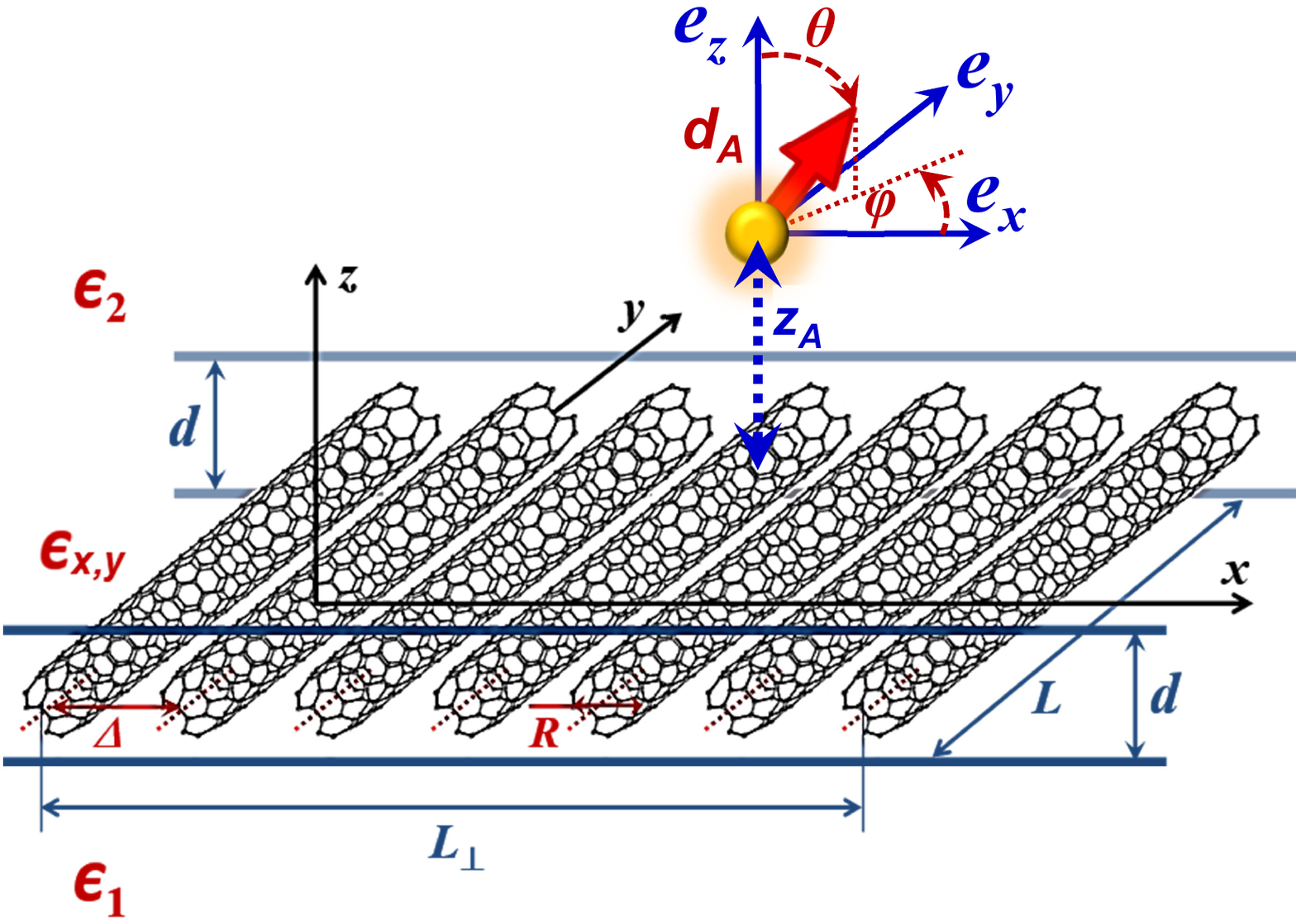}
\caption{Transition dipole orientation for an atomic two-level dipole emitter close to a plane-parallel periodically aligned array of SWCNs of radius $R$ and length $L$ inside of a dielectric of thickness $d$. The array parameters are assumed to obey the set of constraints $2R\!\le\!d\!\ll\!L\!\sim\!L_\perp$ with $R$ and $\Delta$ being much less than the wavelength of light radiation.}
\label{fig1}
\end{figure}

In this work, our focus is on spontaneous emission and fluorescence of atoms (or molecules) in close proximity to a finite-thickness ultrathin film made of an array of periodically aligned densely-packed SWCNs as sketched in Fig.~\ref{fig1} with parameters indicated in the caption. Such an ultra\-thin SWCN metasurface is assumed to be in the TD regime with its in-plane-anisotropic EM response being effectively 2D while still retaining thickness to represent the finite out-of-plane size of the system~\cite{Bondarev2021}. Due to the SWCN chirality, diameter and composition variation, such quantum systems can be used to design single-photon sources with extraordinary adjustable polarization properties for quantum optics applications~\cite{Bond2022,Bay2000}, or more generally for quantum nanophotonics~\cite{Sentef2022,GVidal2024}, to develop a new generation of miniaturized versatile nano\-devices such as highly-selective optical and infrared sensors for single atom (molecule) detection, trapping and manipulation~\cite{Bondarev2015a,Bondarev2015}, including even molecular chemical reactivity control~\cite{Jeremy2019}. We study theoretically both near-field spontaneous emission and far-field resonance fluorescence by atomic type two-level dipole emitters in these systems, focusing on the anisotropic photon emission enhancement effects. While spontaneous emission rate variation has been previously investigated for a variety of nanosystems, such as single SWCNs~\cite{BondLamb04,BondLamb06}, plasmonic nanomaterials~\cite{Bondarev2020,Jeremy2019}, semiconductor-metal interfaces~\cite{Sadeghi2018,Chang22}, and photonic crystals~\cite{Calajo2017,Hughes2005,Kress2005}, to mention a few, the directionality effects in spontaneous emission, resonance fluorescence and scattering of light by quantum dipole emitters near in-plane anisotropic TD metasurfaces still require proper theoretical attention.

We use the fully quantized medium-assisted Quantum Electro\-dynamics (QED) approach~\cite{WelschQO}, with spontaneous emission and fluorescence generated by an excited quantum two-level dipole emitter in close proximity to an ultra\-thin closely-packed SWCN film, a particular case of the in-plane anisotropic TD optical metasurface. In such systems, the confinement-induced EM response nonlocality is known to play a crucial role and no semiclassical theory of EM wave propagation is expected to work~\cite{Buhmann2012}. TD materials restructure the spectral and spatial distribution of EM modes pertaining to not only real but also virtual vacuum processes such as spontaneous emission and van der Waals coupling~\cite{Bondarev2020,Bondarev2023}. While the real modes can be described semiclassically, the physical consequences of the vacuum restructuring with new EM modes generated spontaneously can only be understood in terms of a fully quantized medium-assisted QED formalism~\cite{Andrews2016,Ginzburg2016}. Semiclassical results are only identical to those of QED for processes like single-photon absorption where no modes are spontaneously generated in the process~\cite{Andrews2016}. Converse examples are dipolar spontaneous emission enhancement due to the dipole emitter coupling to the TD epsilon-near-zero modes~\cite{Bondarev2020,Bond2022} and interatomic Coulomb decay~\cite{Buhmann2018}, studied recently within the framework of medium-assisted QED, to yield emission rates orders of magnitude greater than those expected semiclassically.

We show that the SWCN periodic parallel alignment provides an extra measure for the spontaneous emission and resonance fluorescence control of the quantum dipole emitter near the SWCN metasurface---in addition to the metasurface thickness and composition parameters such as SWCN diameter, chirality and translational period. The dipolar spontaneous emission rate and photon fluorescence intensity exhibit highly anisotropic behavior, being enhanced by orders of magnitude in the plane perpendicular to the SWCN alignment along the direction perpendicular to the metasurface plane, in contrast to the commonly believed viewpoint of their uncontrollably random directionality. We emphasize and give a particular example to demonstrate that, contrary to metallic nano\-structures such as plane-periodic plasmonic nanoantenna arrays operating predominantly in the near-IR~\cite{Chen20}, semiconducting SWCN arrays exhibit their remarkable properties at frequencies adjustable broadly from IR through visible range by means of SWCN diameter, chirality, composition, and charge density variation~\cite{Roberts2020,Bondarev2021}. The following sections describe our model, present our theoretical results, and conclude our work by summarizing its key findings. The details of the most cumbersome calculations are presented in the two appendices at the end.

\section{Medium-Assisted QED Formalism for Transdimensional Metasurfaces}

The model system we study here is presented in Fig.~\ref{fig1}. A~horizontal array of parallel aligned ($y$-direction) identical SWCNs of radius $R$ and length $L$ has the translational unit $\Delta$ and width $L_\perp$ ($x$-direction). The array is embedded in a solid dielectric of thickness $d$ to form a composite layer with effective static dielectric permittivity $\epsilon$. This layer is sandwiched between the substrate and superstrate with dielectric permittivities $\epsilon_1$ and $\epsilon_2$, respectively. The structural parameters of the array obey the set of constraints as follows $2R\!\le\!\Delta\!\le\!d\!\ll\!L\!\sim\!L_\perp$ with $R$ and $\Delta$ being much less than the wavelength of the light radiation. The SWCN of the $(m,n)$ type ($n\!\le\!m$) has the radius $R\!=(\!\sqrt{3}\,b/2\pi)\sqrt{m^2+mn+n^2}$, where $b\!=\!1.42$~\AA\space is the C$\,$-C interatomic distance, with the electron charge density constrained by cylindrical symmetry to be uniformly distributed over its surface. The positive background of nuclei keeps the entire system electrically neutral. We consider the most interesting case of $\epsilon\!\gg\epsilon_1+\epsilon_2$. In this case, with $d$ decreasing and getting less than the in-plane distance between a pair of electrons in the composite layer, due to the strong vertical confinement their Coulomb interaction is mostly contributed by the region outside of the layer. This brings our system, the ultrathin SWCN metasurface, in the TD regime where the dimensionality of the system is reduced from 3D to 2D and the only remnant of the $z$-direction is the layer thickness $d$ to represent the size of the vertical confinement~\cite{Keldysh}.

In the medium-assisted QED approach we use here to describe such TD aligned SWCN metasurfaces, intrinsic quasiparticle (phonon, plasmon, exciton, etc.) relaxation phenomena started by absorption of light are considered to create random field fluctuations superposed on those of the physical vacuum where no medium is present. These medium-assisted fluctuating vacuum-type EM fields can be represented by the quantum electric and magnetic field operators (Schr\"{o}dinger picture, Gaussian units)
\begin{eqnarray}
\hat{\mathbf{E}}(\mathbf{r})=\hat{\mathbf{E}}^{(+)}(\mathbf{r})+\hat{\mathbf{E}}^{(-)}(\mathbf{r}),\hskip1.7cm\nonumber\\[-0.3cm]
\label{Ew}\\[-0.1cm]
\hat{\mathbf{E}}^{(+)}(\mathbf{r})=\int_{0}^{\infty}\underline{\hat{\mathbf{E}}}(\mathbf{r},\omega)\;d\omega\,,\;\;\;
\hat{\mathbf{E}}^{(-)}(\mathbf{r})=[\hat{\mathbf{E}}^{(+)}(\mathbf{r})]^{\dag},\nonumber\\
\hat{\mathbf{H}}(\mathbf{r})=-i\frac{c}{\omega}\bm{\nabla}\times\hat{\mathbf{E}}(\mathbf{r}).\hskip2cm
\end{eqnarray}
Their respective Fourier-image vector components are ($\textbf{r}=\bm{\rho}+\!z\textbf{e}_z$, $\alpha=x,y,z$)
\begin{eqnarray}
\underline{\hat{E}\!}_{\,\alpha}(\textbf{r},\omega)=i\frac{4\pi\omega}{c^2}\!\int\!d\bm{\rho}^\prime\!\!\!\!
\sum_{\lambda=x,y,z}\!\!\!\!G_{\alpha\lambda}(\textbf{r},\bm{\rho}^\prime,\omega)\,\underline{\hat{J}\!}_{\,\lambda}(\bm{\rho}^\prime,\omega)\label{Maxwell}\\
\underline{\hat{J}\!}_{\,\alpha}(\bm{\rho},\omega)=
\frac{\omega}{2\pi}\sqrt{\hbar d\,\mbox{Im}\,\varepsilon_{\alpha\alpha}(\bm{\rho},\omega)}\,\hat{f}_\alpha(\bm{\rho},\omega),\hskip0.4cm\label{noise}\\
\underline{\hat{H}\!}_{\,\alpha}(\textbf{r},\omega)=
-i\frac{c}{\omega}\!\sum_{\lambda,\mu=x,y,z}\!\!\!\!e_{\alpha\lambda\mu}\nabla_{\!\lambda}\,\underline{\hat{E}\!}_{\,\mu}(\textbf{r},\omega).\hskip0.55cm
\end{eqnarray}
Medium composition is included in this QED equation set by means of the imaginary part of the EM response tensor $\varepsilon_{\alpha\alpha}(\bm{\rho},\omega)$. This tensor is assumed to be diagonal and independent of the vertical $z$-coordinate, which for ultrathin optically dense TD films is represented by thickness $d$ of the film~\cite{Bondarev2017a,Bondarev2018,Bondarev2020}. Medium confinement geometry is defined by the classical EM field Green's tensor $G_{\alpha\lambda}$, which can be found for a confined material system of interest under appropriate boundary conditions and radiation conditions at infinity~\cite{WelschQO,Tomas95}. The operator $\underline{\hat{J}\!}_{\,\alpha}$ is the Fourier-image of the noise current density operator responsible for medium absorption whereby single-quantum vacuum-type medium excitations are created (annihilated) by bosonic operators $\hat{f}^\dagger_\alpha$ ($\hat{f}_\alpha$) such that
\begin{equation}
\big[\hat{f}_\alpha(\bm{\rho},\omega),\hat{f}^\dagger_\beta(\bm{\rho}^\prime,\omega^\prime)\big]\!
=\delta_{\alpha\beta}\delta(\bm{\rho}-\bm{\rho}^\prime)\delta(\omega-\omega^\prime).
\label{fcommut}
\end{equation}
The presence of the noise current density operator is consistent with the Fluctuation-Dissipation Theorem and is necessary to warrant the correct equal-time electric and magnetic field operator commutation relations of vacuum QED in the presence of medium absorption~\cite{WelschQO}.

\section{Dipole emitter and metasurface: the hamiltonian}

In terms of the medium-assisted QED scheme above, the following second-quantized Hamiltonian describes the coupled atom-field system with an atom (or a molecule) at an arbitrary point $\mathbf{r}_{A}=\bm{\rho}_A+z_A\textbf{e}_z$ above the SWCN film (as sketched in Fig.~\ref{fig1}) within the framework of the electric dipole and two-level approximations~\cite{WelschQO,BondLamb06}
\begin{eqnarray}
\hat{H}=\!\int_{0}^{\infty}\!\!\!\!\!d\omega\!\int\!d\bm{\rho}\!\!\!\sum_{\lambda=x,y,z}\!\!\!\hbar\omega\hat{f}^{\dag}_\lambda(\bm{\rho},\omega)\hat{f}_\lambda(\bm{\rho},\omega)\hskip0.75cm\nonumber\\[-0.5cm]
\label{Ham}\\
+{1\over{2}}\,\hbar\omega_{A}\,\hat{\sigma}_{z}-\!\!\!\sum_{\lambda=x,y,z}\!\!\!d_{\lambda}\Big[\hat{E}^{(+)}_{\lambda}(\mathbf{r}_{A})\hat{\sigma}^{\dag}+\hat{E}^{(-)}_{\lambda}(\mathbf{r}_{A})\hat{\sigma}\Big]\nonumber
\end{eqnarray}
Here, the three terms represent the quantum medium-assisted field subsystem, the atomic subsystem modeled by a quantum two-level dipole emitter with frequency $\omega_A$, and their interaction, respectively. The Pauli operators $\hat{\sigma}_{z}\!=\!|u\rangle\langle u|\!-\!|l\rangle\langle l|$, $\hat{\sigma}\!=\!|l\rangle\langle u|$, $\hat{\sigma}^{\dag}\!=\!|u\rangle\langle l|$ describe the transitions in the atomic subsystem between its upper $|u\rangle$ and lower $|l\rangle$ states. The interaction term is due to the coupling of the atomic transition dipole $d_\lambda\!=\!\langle u|\hat{d}_\lambda|l\rangle$ ($\lambda\!=\!x,y,z$) to the quantum medium-assisted electric field of Eqs.~(\ref{Ew})--(\ref{noise}) where the medium composition is represented by the imaginary part of the EM response tensor of the SWCN film. The frequency $\omega_{A}$ is generally red-shifted relative to the transition dipole frequency $\omega_{ul}$, so that $\omega_A\!=\omega_{ul}+\delta\omega_{ul}$ ($\delta\omega_{ul}\!<\!0$), due to the van der Waals type coupling of the dipole emitter to the film~\cite{BondLamb06,BondLamb05}. With no loss of generality, the shift $\delta\omega_{ul}$ will be treated as being included in $\omega_A$ in what follows. Its general properties can be found in Ref.~\cite{WelschQO}.

Collective EM response of the TD films made of periodically aligned SWCN arrays was recently studied theoretically~\cite{Bondarev2019,Bondarev2021}. With contributions from both plasmons and excitons (corresponding to intra- and interband transitions in the far-infrared and optical spectral regions, respectively), it was shown to be strongly anisotropically nonlocal due to the cylindrical spatial anisotropy, periodic in-plane transverse inhomogeneity, and vertical quantum confinement of the system. In the reciprocal (momentum) space the anisotropic EM response of such an ultrathin finite-thickness metasurface can be represented by the 3D tensor
\begin{equation}
\hat{\varepsilon}=\left[\begin{array}{ccc}
\epsilon_x & 0 & 0 \\
0 & \varepsilon(q,\omega) & 0 \\
0 & 0 & \epsilon_z\\
\end{array}\right]=\varepsilon_\mu\delta_{\mu\lambda}\;\;\;(\mu,\lambda=x,y,z).
\label{EMtensor}
\end{equation}
Here, $\epsilon_x\!=\!\epsilon$ is the effective constant permittivity of the dielectric layer with the aligned SWCN array immersed in it. This is the response of the metasurface in the $x$-direction perpendicular to the SWCN alignment as shown in Fig.~\ref{fig1}. The response of the ultrathin metasurface structure in the $z$-direction makes no effect on light propagation. We take it to be $\epsilon_z\!=\!(\epsilon_1+\epsilon_2)/2$ in what follows since the range of thicknesses does not exceed the wave length of light in the TD regime and the transverse polarizability of SWCNs is negligible. The function
\begin{equation}
\varepsilon(q,\omega)=\epsilon-\frac{2\epsilon f_{\rm CN}\sigma_{yy}(q,\omega)}{f_{\rm CN}\sigma_{yy}(q,\omega)\!+\!i\omega e^2N_{\rm 2D}R/m^\ast\omega^2_p(q)d}
\label{epsilonyy}
\end{equation}
is the nonlocal response of the TD film in the direction of SWCN alignment ($y$-direction). Here $q$, $f_{\rm CN}\!=\!\pi R^2/\Delta d$, and $\sigma_{yy}$ are the $y$-direction quasi\-particle momentum absolute value, the volume fraction of SWCNs, and the complex-valued (to include quasiparticle relaxation) axial surface conductivity of an individual SWCN, respectively. The quantity
\begin{equation}
\omega_p(q)=\sqrt{\frac{4\pi e^2N_{\rm 2D}}{m^{\ast}d}\frac{2qRI_0(qR)K_0(qR)}{\epsilon+(\epsilon_1+\epsilon_2)/(qd)}\,}
\label{plasmaFy}
\end{equation}
stands for the intraband plasma oscillation frequency for a general finite-thickness, cylindrically anisotropic, periodically aligned (metallic or semiconducting) array~\cite{Bondarev2019}. Here, $m^\ast$ is the electron effective mass, $N_{\rm 2D}$~$(=\!N_{\rm 3D}d)$ is the surface electron density, $I_0$ and $K_0$ are the zeroth-order modified cylindrical Bessel functions. The latter are responsible for the correct normalization of the electron density distribution over cylindrical surfaces, to give the isotropic TD film plasma frequency studied previously in Refs.~\cite{Bondarev2017a,Shah2022} in the limit $R\!\rightarrow\!\infty$ (in which case $qRI_0(qR)K_0(qR)\!\rightarrow\!1/2$). More details about the collective EM response of the TD films made of periodically aligned SWCN arrays can be found in Refs.~\cite{Bondarev2019,Bondarev2021}.

\section{Anisotropic Spontaneous Emission}

When an atomic dipole emitter is initially in the upper state and the field subsystem is in vacuum, the time-dependent wave function of the whole system can be written as
\begin{eqnarray}
|\psi(t)\rangle=e^{-i(\omega_{A}/2)t}C_{u}(t)|u\rangle|\{0\}\rangle\hskip2.0cm\nonumber\\[-0.15cm]
\label{wfunc}\\[-0.15cm]
+\!\int_{0}^{\infty}\!\!\!\!\!d\omega\!\int\!d\bm{\rho}\!\!\!\sum_{\lambda=x,y,z}\!\!\!\!e^{-i(\omega-\omega_{A}/2)t}
C_{l\lambda}(\bm{\rho},\omega,t)|l\rangle\hat{f}^\dagger_\lambda(\bm{\rho},\omega)|\{0\}\rangle.\nonumber
\end{eqnarray}
Here, $|\{0\}\rangle$ is the vacuum state of the medium-assisted field subsystem, $\hat{f}^\dagger_\lambda(\bm{\rho},\omega)|\{0\}\rangle$ is its excited state where the field is in a single-quantum Fock state with dipole emitter oriented in the $\lambda\,(=\!\!x,y,z)$-direction, $C_{u}$ and $C_{l\lambda}$ are the population probability amplitudes of the upper state and lower state of the coupled dipole emitter-metasurface system such that
\[
|C_{u}(t)|^2+\!\int_{0}^{\infty}\!\!\!\!\!d\omega\!\int\!d\bm{\rho}\!\!\!\sum_{\lambda=x,y,z}\!\!\!|C_{l\lambda}(\bm{\rho},\omega,t)|^2=1.
\]

Plugging the wave function~(\ref{wfunc}) in the time-dependent Schr\"{o}dinger equation with the Hamiltonian~(\ref{Ham}), one obtains the following coupled equation set for the unknown population probability amplitudes
\begin{eqnarray}
\mbox{\it\.{C}}_{u}(t)=-\frac{2}{\hbar c^2}\!\int_{0}^{\infty}\!\!\!\!\!d\omega\!\int\!d\bm{\rho}\!\!\!\!\sum_{\mu,\lambda=x,y,z}\!\!\!\!\omega^2e^{-i(\omega-\omega_{A})t}\hskip0.3cm\nonumber\\[-0.3cm]
\label{popampu}\\[-0.2cm]
\times\sqrt{\hbar d\,\mbox{Im}\,\varepsilon_{\lambda\lambda}(\bm{\rho},\omega)}\,d_{\mu}G_{\mu\lambda}(\textbf{r}_{A},\bm{\rho},\omega)C_{l\lambda}(\bm{\rho},\omega,t),\nonumber\\[0.2cm]
\mbox{\it\.{C}}_{l\lambda}(\bm{\rho},\omega,t)=\frac{2}{\hbar c^2}\!\!\!\sum_{\mu=x,y,z}\!\!\!\omega^2e^{i(\omega-\omega_{A})t}\hskip1.0cm\nonumber\\[-0.3cm]
\label{popampl}\\[-0.2cm]
\times\sqrt{\hbar d\,\mbox{Im}\,\varepsilon_{\lambda\lambda}(\bm{\rho},\omega)}\,d_{\mu}G_{\mu\lambda}^\ast(\textbf{r}_{A},\bm{\rho},\omega)C_{u}(t).\hskip0.5cm\nonumber
\end{eqnarray}
Substituting the result of the integration of Eq.~(\ref{popampl}) [see Eq.~(\ref{A3})] into Eq.~(\ref{popampu}) and using the identity
\begin{eqnarray}
\frac{\omega^2}{c^2}\!\int\!\!d\bm{\rho}\!\!\!\sum_{\lambda=x,y,z}\!\!\!\!d\,\mbox{Im}\,
\varepsilon_{\lambda\lambda}(\bm{\rho},\omega)G_{\alpha\lambda}(\textbf{r},\bm{\rho},\omega)G^\ast_{\beta\lambda}(\textbf{r}^\prime\!,\bm{\rho},\omega)\nonumber\\[-0.25cm]
\label{mainidentity}\\[-0.25cm]
=\mbox{Im}\,G_{\alpha\beta}(\textbf{r},\textbf{r}^\prime\!,\omega)\hskip2.5cm\nonumber
\end{eqnarray}
(a particular case of the general 3D Green tensor integral relation, see Ref.~\cite{WelschQO}), with initial conditions $C_{u}(0)\!=\!1$ and $C_{l\lambda}(\bm{\rho},\omega,0)\!=\!0$ it is finally straightforward to obtain
\begin{equation}
C_{u}(t)=1+\int_{0}^{t}\!\!\!d\tau K(t-\tau)\,C_{u}(\tau).
\label{Volterra}
\end{equation}
The kernel of this integral (Volterra) equation is
\begin{equation}
K(t)=\frac{1}{2\pi}\!\int_{0}^\infty\!\!\!\!\!d\omega\frac{e^{-i(\omega-\omega_{A})t}-1}{i(\omega-\omega_{A})}\Gamma(\textbf{r}_{A},\omega)
\label{kernel}
\end{equation}
with
\begin{equation}
\Gamma(\textbf{r}_{A},\omega)=\frac{8\pi\omega^{2}}{\hbar c^{2}}\!\!\!\!\!\sum_{\mu,\lambda=x,y,z}\!\!\!\!\!d_\mu d_\lambda\,\mathrm{Im\,}G_{\mu\lambda}(\textbf{r}_{A},\textbf{r}_{A},\omega),
\label{Gammaraomega}
\end{equation}
the dipole spontaneous emission rate as a function of~$\omega$, written in terms of the imaginary part of the equal-position EM field Green tensor---a particular case of the tensor $\mbox{Im}\,G_{\mu\lambda}(\textbf{r},\textbf{r}_A,\omega)$. The latter can be split into two parts
\begin{equation}
\mbox{Im\,}G_{\mu\lambda}(\textbf{r},\textbf{r}_A,\omega)\!=
\!\mathrm{Im}\big[G^{\mathrm{\,0}}_{\mu\lambda}(\textbf{r},\textbf{r}_{A},\omega)\!+\!G^{\mathrm{\,sc}}_{\mu\lambda}(\textbf{r},\textbf{r}_{A},\omega)\big]\hskip-0.15cm
\label{split}
\end{equation}
to represent the free space and metasurface scattering contributions, respectively.

The free space contribution of Eq.~(\ref{split}) comes out of the well-known EM field Green tensor of the free dielectric space,
\begin{equation}
G^{\mathrm{\,0}}_{\mu\lambda}(\textbf{r},\textbf{r}_{A},\omega)=
\left(\frac{1}{\kappa^2}\nabla_\mu\nabla_\lambda+\delta_{\mu\lambda}\right)\frac{e^{i\kappa|\textbf{r}-\textbf{r}_A|}}{4\pi|\textbf{r}-\textbf{r}_A|},
\label{Greenfree}
\end{equation}
where $\kappa\!=\!k_0\sqrt{\epsilon_2}$ in our case and $k_0\!=\!\omega/c$. Here, it is straight\-forward to perform the differentiation, followed by the 3D space averaging $(1/4\pi)\!\int\!d\Omega$, to obtain
\begin{equation}
\mathrm{Im\,}G^{\mathrm{\,0}}_{\mu\lambda}(\textbf{r},\textbf{r}_{A},\omega)=\frac{\sin(\kappa|\textbf{r}-\textbf{r}_{A}|)}{6\pi|\textbf{r}-\textbf{r}_{A}|}\delta_{\mu\lambda},
\label{ImGreenfree}
\end{equation}
whereby
\begin{equation}
\mathrm{Im\,}G_{\mu\lambda}(\textbf{r}_{A},\textbf{r}_{A},\omega)=
\frac{\kappa}{6\pi}\delta_{\mu\lambda}+\mathrm{Im\,}G^{\mathrm{\,sc}}_{\mu\lambda}(\textbf{r}_{A},\textbf{r}_{A},\omega).
\label{Gscatt}
\end{equation}
With this taken into account, Eq.~(\ref{Gammaraomega}) can be rewritten in the principle axes system of the $G^{\mathrm{\,sc}}_{\mu\lambda}$ tensor, which is dictated by the EM response tensor (\ref{EMtensor}) of the metasurface, as follows
\begin{equation}
\Gamma(\textbf{r}_{A},\omega)=\Gamma_0(\omega)\Big[1+\xi(\textbf{r}_{A},\omega)\Big].
\label{Gamma0}
\end{equation}
Here, $\Gamma_0(\omega)\!=\!4\omega^3\texttt{d}^2\sqrt{\epsilon_2}/(3\hbar c^3)$ is the isotropic dipolar spontaneous emission rate in the free dielectric space, with $\texttt{d}^2\!=\!\sum_\mu d_\mu^2$ and $d_\mu\!=\!\texttt{d}(\sin\vartheta\cos\varphi,\sin\vartheta\sin\varphi,\cos\vartheta)$ as indicated in Fig.~\ref{fig1}. The function
\begin{equation}
\xi(\textbf{r}_{A},\omega,\vartheta,\varphi)=\frac{6\pi}{\kappa\texttt{d}^2}\!\!\!\sum_{\mu=x,y,z}\!\!\!d_\mu^2\,\mathrm{Im\,}G^{\mathrm{\,sc}}_{\mu\mu}(\textbf{r}_{A},\textbf{r}_{A},\omega)
\label{PDOS}
\end{equation}
is the anisotropic local photonic density of states (LDOS) that includes the metasurface presence effect relative to vacuum.

\section{Resonance Fluorescence}

In terms of our medium-assisted QED approach, the fluorescence intensity of an atomic dipole emitter near the SWCN metasurface can be written as
\begin{equation}
I(\textbf{r},t)=\!\!\!\sum_{\mu=x,y,z}\!\!\!\langle\psi(t)|\hat{\sigma}\hat{E}^{(-)}_{\mu}(\mathbf{r})\hat{E}^{(+)}_{\mu}(\mathbf{r})\hat{\sigma}^{\dag}|\psi(t)\rangle,
\label{Intensity}
\end{equation}
where the field operators are defined by Eqs.~(\ref{Ew})--(\ref{noise}) with the metasurface EM response tensor of Eq.~(\ref{EMtensor}) and the wave function of the system given by Eq.~(\ref{wfunc}). From here, using Eq.~(\ref{mainidentity}) and the coefficients $C_{l\lambda}$ obtained by integration of Eq.~(\ref{popampl}), it is possible to bring the intensity of interest to the final form as follows (see Appendix A)
\begin{equation}
I(\textbf{r},t)=\!\!\!\sum_{\mu=x,y,z}\left|\int_{0}^{t}\!\!\!d\tau K_\mu(t-\tau)C_u(\tau)e^{-i\omega_{A}\tau}\right|^2,
\label{IntensityK}
\end{equation}
where
\begin{equation}
K_\mu(t)=\frac{4}{c^2}\!\int_{0}^\infty\!\!\!\!\!d\omega\,\omega^2e^{-i\omega t}\!\!\!\!\sum_{\lambda=x,y,z}\!\!\!\mbox{Im}\,G_{\mu\lambda}(\textbf{r},\textbf{r}_A,\omega)d_\lambda\,.
\label{Klambda}
\end{equation}
In view of Eqs.~(\ref{split}) and (\ref{ImGreenfree}) this can also be brought to the form
\begin{eqnarray}
K_\mu(t)=\frac{2\texttt{d}}{3\pi c^3}\!\int_{0}^\infty\!\!\!\!\!d\omega\,\omega^3e^{-i\omega t}\xi_\mu(\textbf{r},\textbf{r}_A,\omega)\nonumber\\[-0.15cm]
\label{Kmuxi}\\[-0.15cm]
\times\!\left[1+\frac{\sin(\kappa|\textbf{r}-\textbf{r}_{A}|)}{6\pi|\textbf{r}-\textbf{r}_{A}|\,\xi_\mu(\textbf{r},\textbf{r}_A,\omega)}\right],\hskip0.7cm\nonumber
\end{eqnarray}
where
\begin{equation}
\xi_\mu(\textbf{r},\textbf{r}_A,\omega,\vartheta,\varphi)=\frac{6\pi d_\mu}{\kappa\texttt{d}}\,\mathrm{Im\,}G^{\mathrm{\,sc}}_{\mu\mu}(\textbf{r},\textbf{r}_A,\omega)
\label{xirar}
\end{equation}
are the anisotropic distance-dependent photonic LDOS functions related to the LDOS of Eq.~(\ref{PDOS}) as follows
\[
\sum_{\mu=x,y,z}\!\!\!\!\frac{d_\mu}{\texttt{d}}\,\xi_\mu(\textbf{r},\textbf{r}_A,\omega,\vartheta,\varphi)\Big|_{\textbf{r}=\textbf{r}_A}\!\!\!=\xi(\textbf{r}_A,\omega,\vartheta,\varphi).
\]

Resonance fluorescence results from $\omega_A$ being in resonance with a peak frequency $\omega_r$ of the LDOS in Eq.~(\ref{Gamma0}). With the latter approximated by the Lorentzian function $\xi(\textbf{r}_A,\omega)\!\approx\!\xi(\textbf{r}_A,\omega_{r})\delta\omega_{r}^{2}/[(\omega-\omega_{r})^{2}+\delta\omega_{r}^{2}]$ of half-width-at-half maximum $\delta\omega_r$, where $\xi(\textbf{r}_A,\omega_{r})\gg1$, Eq.~(\ref{Volterra}) can be solved analytically to yield $C_u(\tau)$ in Eq.~(\ref{IntensityK}) in the following explicit form
\begin{equation}
C_{u}(\tau)\approx c_{+}e^{-{\displaystyle c}_{-}\delta\omega_r\tau/({\displaystyle c}_{-\!}-{\displaystyle c}_{+})}\!
+c_{-}e^{-{\displaystyle c}_{+}\delta\omega_r\tau/({\displaystyle c}_{+\!}-{\displaystyle c}_{-})}\!.\hskip-0.1cm
\label{singleres}
\end{equation}
Here, $c_{\pm}\!=\!\big[1\pm1/\sqrt{1-(\Omega/\delta\omega_r)^2}\,\big]/2$ and the Rabi frequency $\Omega\!=\!\sqrt{2\delta\omega_r\Gamma(\textbf{r}_A,\omega_A\!\approx\!\omega_r)}$ is to represent the dipole emitter level hybridization due to the coupling to the quantum medium-assisted modes of the material subsystem (excitons and plasmons in the parallel aligned SWCNs of our TD metasurface). The coupling is termed weak if $(\Omega/\delta\omega_r)^2\!\ll\!1$, in which case Eq.~(\ref{singleres}) yields the fast exponential decay time dynamics $C_u(\tau)\!\sim\!\exp[-\Gamma\tau/2]$ for the upper dipole emitter state, or strong if $(\Omega/\delta\omega_r)^2\!\gg\!1$, in which case it yields slow decay dynamics with Rabi oscillations $C_u(\tau)\!\sim\!\exp[-\delta\omega_r\tau/2]\cos(\Omega\tau/2)$.

Using Eq.~(\ref{singleres}) inside of Eq.~(\ref{Kmuxi}) and observing that the second (oscillatory) term in square brackets of Eq.~(\ref{Kmuxi}) is negligible for $|\textbf{r}-\textbf{r}_{A}|\!\gg\!|\textbf{r}_{A}|$ and so can be dropped, the following expression can be obtained for sufficiently long time in this most relevant limit (see Appendix B)
\begin{eqnarray}
I(\textbf{r},t)\!\approx\!\Big(\frac{\texttt{2d}}{3c^3}\Big)^{\!2}\!\!\!\!\sum_{\mu=x,y,z}\!\!\Big|c_{+}\xi_\mu(\textbf{r},\textbf{r}_A,\omega_{-})\omega_{-}^3e^{-i\omega_{-}t}\nonumber\\[-0.25cm]
\label{Ifin}\\[-0.25cm]
+\,c_{-}\xi_\mu(\textbf{r},\textbf{r}_A,\omega_{+})\omega_{+}^3e^{-i\omega_{+}t}\Big|^2\hskip1.25cm\nonumber
\end{eqnarray}
with $\omega_\pm\!=\omega_A-ic_\pm\delta\omega_r/(c_\pm\!-c_\mp)$. This asymptotic long-distance, long-time expression can be simplified further to give the weak and strong dipole emitter-metasurface coupling cases individually. Using the first-nonvanishing-order Taylor series expansions for the $c_{\pm}$ coefficients in it leads to
\begin{equation}
I_w(\textbf{r},t)\approx I_0(\textbf{r},\omega_A)\,e^{-\Gamma(\textbf{r}_{A},\omega_A)t}
\label{Ifinweak}
\end{equation}
and
\begin{equation}
I_s(\textbf{r},t)\approx I_0(\textbf{r},\omega_A)\cos^2\!\!\Big(\frac{\Omega t}{2}\Big) e^{-\delta\omega_rt}
\label{Ifinstrong}\\
\end{equation}
for weak and strong dipole emitter-metasurface coupling, respectively, with
\begin{equation}
I_0(\textbf{r},\omega_A)=\Big(\frac{\texttt{2d}}{3c^3}\Big)^{\!2}\!\!\!\!\sum_{\mu=x,y,z}\Big|\xi_\mu(\textbf{r},\textbf{r}_A,\omega_A)\Big|^2\!,
\label{xiws}
\end{equation}
where $\omega_A\!\approx\omega_r$. Thus, the fluorescence intensity time dynamics (fast exponential for the former and slow oscillatory for the latter) are both controlled by the largely increased anisotropic amplitude (distance-dependent photonic LDOS squared) due to the emitter-metasurface interaction.

\section{Results and Discussion}

As can be seen from the above, the photonic LDOS of Eq.~(\ref{PDOS}) and its distance-dependent analogue of Eq.~(\ref{xirar}) are the key quantities to control the near-field EM processes such as spontaneous emission and resonance fluorescence, respectively. Both of them are determined by the scattering part of the EM field Green tensor of a planar multilayer structure, which can be diagonalized as per the in-plane layer symmetry and can generally be factored into the in-plane and out-of-plane components~\cite{Tomas95}. The distance dependence of such a tensor is largest (and so of the most interest) in the out-of-plane direction along the $z$-axis (see Fig.~\ref{fig1}), where it takes the form
\begin{eqnarray}
G^{\mathrm{\,sc}}_{\mu\mu}(z,z_A,\omega)=G^{\mathrm{\,sc}}_{\mu\mu}(\bm{\rho},\bm{\rho}_A;z,z_A,\omega)\Big|_{\bm{\rho}=\bm{\rho}_A}\!\!\!=\nonumber\\[-0.1cm]
\label{GFofinterest}\\[-0.1cm]
\int\!\!d\textbf{k}\,e^{i\textbf{k}\cdot(\bm{\rho}-\bm{\rho}_A)}G^{\mathrm{\,sc}}_{\mu\mu}(\textbf{k},k_\perp,\omega)\Big|_{\bm{\rho}=\bm{\rho}_A}
e^{ik_\perp(z-z_A)}.\nonumber
\end{eqnarray}
Here, $\textbf{k}\!=\!k_x\textbf{e}_x+k_y\textbf{e}_y\!=\!k_\rho\textbf{e}_\rho$ is the in-plane momentum with absolute value $k\!=\!|\textbf{k}|\!=\!\sqrt{k_x^2+k_y^2}$, and $k_\perp\!=\!|\textbf{k}_\perp|$ is the absolute value of the out-of-plane momentum component $\textbf{k}_\perp\!=\!k_\perp\textbf{e}_z$ in the positive direction of the $z$-axis.

For our anisotropic single-layer case, Eq.~(\ref{GFofinterest}) can be written in terms of the reflection coefficients for spontaneously emitted $s$- and $p$-polarized photons (\emph{TE} and \emph{TM} waves, respectively) as follows~\cite{Tomas95,BuhmannPRA08}
\begin{eqnarray}
G^{\mathrm{\,sc}}_{\mu\mu}(z,z_{A},\omega)=\!\frac{i}{2}\int^{+\infty}_{0}\!\!\!\!\!\!\!dk\frac{k}{\beta_2}\,R_{\mu\mu}(k)\,e^{i\beta_{2}(z+z_{A})},\hskip0.75cm\label{Greentensor}\\
R_{xx,yy}=r^{s[x,y]}_{2-}(k)-\frac{\beta^{2}_{2}c^{2}}{\varepsilon_{2}\omega^{2}}\,r^{p[x,y]}_{2-}(k),\hskip1.3cm\nonumber\\
R_{zz}=\frac{2k^{2}c^{2}}{\varepsilon_{2}\omega^{2}}\,r^{p[z]}_{2-}(k),\hskip2.75cm\nonumber
\end{eqnarray}
where the reflection coefficients in standard notations are
\begin{eqnarray}
r^{\sigma[\mu]}_{2-}=\frac{r^{\sigma[\mu]}_{2}-r^{\sigma[\mu]}_{1}e^{2i\beta_\mu d}}{1-r^{\sigma[\mu]}_{1}r^{\sigma[\mu]}_{2}e^{2i\beta_\mu d}}\;\;(\sigma=s,p;\;\mu=x,y,z),\hskip0.5cm\label{rsigmaxyz}\\
r^{s[x,y]}_{j}=\frac{\beta_{j}-\beta_{x,y}}{\beta_{j}+\beta_{x,y}}\,,\;\;\;\beta_{j}=\sqrt{\epsilon_{j}k_0^2-k^{2}}\;\;\;(j=1,2),\hskip0.4cm\nonumber\\
r^{p[x,z]}_{j}=\frac{\beta_{j}\epsilon_{x,z}-\beta_{x,z}\epsilon_{j}}{\beta_{j}\epsilon_{x,z}+\beta_{x,z}\epsilon_{j}},\;\;\;
r^{p[y]}_{j}=\frac{\beta_{j}\varepsilon(k,\omega)-\beta_y\epsilon_{j}}{\beta_{j}\varepsilon(k,\omega)+\beta_y\epsilon_{j}},\hskip0.35cm\nonumber\\
\beta_{x,z}=\sqrt{\epsilon_{x,z}k_0^2-k^{2}},\;\;\;\beta_y=\sqrt{\varepsilon(k,\omega)k_0^2-k^{2}}\hskip1.0cm\nonumber
\end{eqnarray}
according to the EM response tensor (\ref{EMtensor}) of our system. Here, $\beta_{1,2}$ and $\beta_{x,y,z}$ are the absolute values of the photon momentum $z$-components in region~$1$ (substrate with dielectric constant $\epsilon_1$), in region~$2$ (dielectric constant $\epsilon_2$) where the dipole emitter is situated, and in the $x,y,z$-directions of the region bounded by the anisotropic metasurface, respectively. Rescaling of the quantities in Eq.~(\ref{Greentensor}) by
\begin{equation}
z+z_A\!=\!\frac{l\!+\!l_{\!A}}{2\kappa},R_{\mu\mu}\!=\!\frac{2\bar{R}_{\mu\mu}}{\kappa},k\!=\!\kappa t,i\beta_2\!=\!\kappa x,\beta_2\!=\!\kappa y\hskip-0.1cm
\label{dimlessunits}
\end{equation}
with $\kappa\!=\!k_0\sqrt{\varepsilon_2}\,$, allows one to rewrite it as a sum of the two well-defined single-valued real integrals of the form
\begin{eqnarray}
G^{\mathrm{\,sc}}_{\mu\mu}(z,z_{A},\omega)=
i\!\!\int^{+\infty}_{0}\!\!\!\!\!dt\,\frac{t\bar{R}_{\mu\mu}(t)}{\sqrt{1\!-\!t^2}}\,e^{i\sqrt{1-t^2}(l+l_{\!A})/2}\label{propevan}\hskip0.5cm\\
=\!\!\int^{+\infty}_{0}\!\!\!\!\!\!\!dx\bar{R}_{\mu\mu}\big(\!\sqrt{1\!+\!x^2}\,\big)e^{-x(l+l_{\!A})/2}\hskip1.5cm\nonumber\\
+\,i\!\!\int^{1}_{0}\!\!\!dy\bar{R}_{\mu\mu}\big(\!\sqrt{1\!-\!y^2}\,\big)e^{iy(l+l_{\!A})/2}.\hskip1.5cm\nonumber
\end{eqnarray}
Here, the first and second integrals are contributed by the evanescent and propagating waves, respectively, as can be seen from the exponential factors of their integrands. It can also be seen from Eqs.~(\ref{Gamma0}), (\ref{PDOS}) and Eqs.~(\ref{xirar}), (\ref{xiws}) that the former is responsible for the spontaneous emission enhancement of the dipole emitter while the latter makes strong resonance fluorescence possible for it as a two-step process ($\sim\!\sum_\mu|\xi_\mu|^2$ therefore) in which most of the photons emitted spontaneously by the dipole emitter get re-emitted to infinity by the entire metasurface structure.

\begin{figure}[t]
\hskip0.5cm\includegraphics[width=\linewidth]{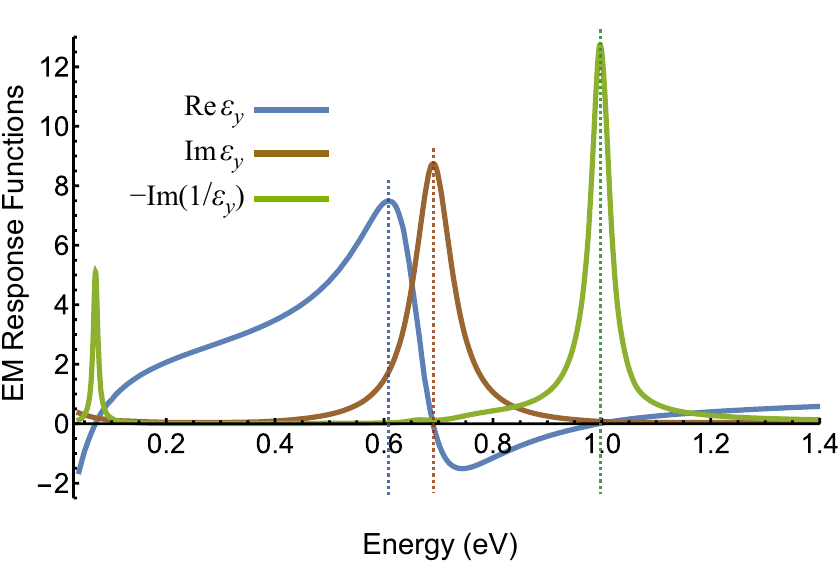}
\caption{The real (blue) and imaginary (brown) parts of the low-energy complex in-plane EM response tensor (\ref{EMtensor}) along the SWCN alignment direction ($y$-direction in Fig.~\ref{fig1}) used as the model EM response for the ultrathin semiconducting SWCN film. The green line shows the plasmonic response. The blue line exhibits the classical and quantum negative refraction bands below $0.1$~eV and between $0.7$~eV and $1$~eV, respectively. The latter is due to the presence of the first exciton absorption resonance (brown line). See text and Ref.~\cite{Bondarev2021} for more details.}\vspace{-0.1cm}
\label{fig2}
\end{figure}

\begin{figure}[t]
\hskip0.5cm\includegraphics[width=\linewidth]{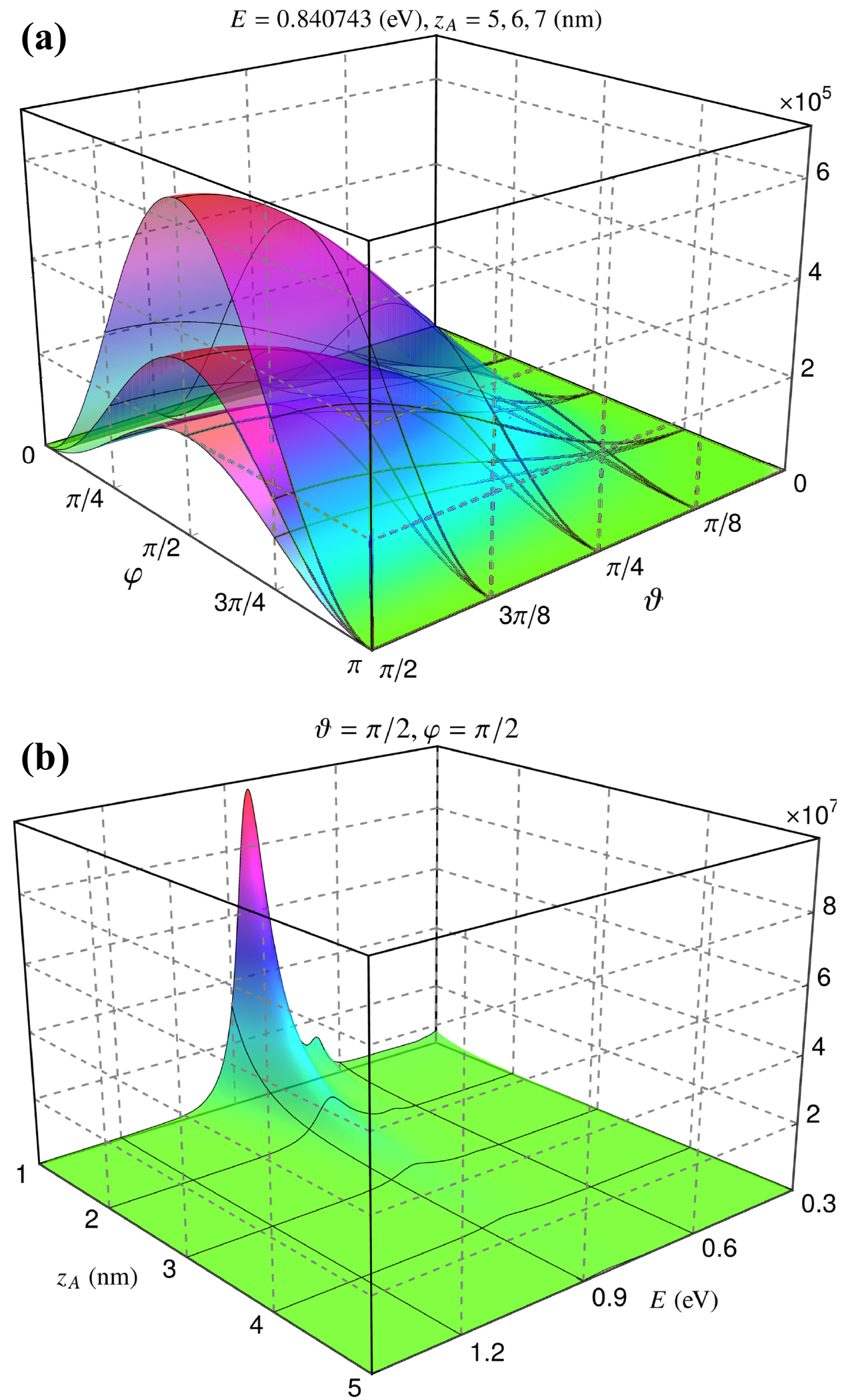}
\caption{(a)~Anisotropic LDOS of Eq.~(\ref{PDOS}) calculated with EM field Green tensor of Eq.~(\ref{Greentensor}) as a function of transition dipole relative orientation angles for three dipole emitter-metasurface distances as sketched in Fig.~\ref{fig1} (decreasing distance increases the LDOS). (b)~Fixed-angle max LDOS of Eq.~(\ref{PDOS}) as a function of dipole emitter-metasurface distance and energy.}\vspace{-0.1cm}
\label{fig3}
\end{figure}

Figure~\ref{fig2} shows a typical example of the real part $\mbox{Re}\,\varepsilon_y$ and imaginary part $\mbox{Im}\,\varepsilon_y$ of the $y$-component of the low-energy in-plane EM response tensor (\ref{EMtensor}) for ultrathin SWCN metasurfaces. This is the component along the SWCN alignment direction ($y$-direction in Fig.~\ref{fig1}), which we use in Eqs.~(\ref{EMtensor})--(\ref{plasmaFy}) to simulate the photonic LDOS of Eq.~(\ref{PDOS}) and its distance-dependent analogue of Eq.~(\ref{xirar}). This function was previously reported in Ref.~\cite{Bondarev2021} to have been obtained numerically from Eqs.~(\ref{epsilonyy}) and (\ref{plasmaFy}) with $\epsilon\!=\!10$ (dielectric layer) and $\epsilon_1\!=\!\epsilon_2\!=\!1$ by using the Maxwell-Garnett method for the $10$~nm thick weakly inhomogeneous TD metasurface. The structure was comprised of a quasi\-periodic mixture of the (16,0), (17,0), (18,0), (19,0) and (20,0) metallic and semiconducting SWCN arrays to make the film composition $1/3$ metallic and $2/3$ semiconducting as is normally the case experimentally~\cite{Roberts2020,Falk2020,Roberts2019,Naik2019,Kono2019,Falk2018,Falk2017}. The energy range was chosen to only include the first exciton resonance and thus to reproduce the real and imaginary parts of the in-plane EM response of weakly inhomogeneous self-assembled SWCN metamaterial reported earlier experimentally~\cite{Roberts2019}. The function $\mbox{Re}\,\varepsilon_y$ (blue line) forms a relatively broad negative refraction band in the vicinity of the first exciton transition (resonance of $\mbox{Im}\,\varepsilon_y$ shown by brown line). The presence of this quantum interband electronic transition makes our ultrathin material system behave as a uniaxial hyper\-bolic metasurface with $\mbox{Re}\,\varepsilon_y$ being negative not only in the classical (Drude-like positive dispersion) domain below $0.1$~eV but also in the quantum (negative dispersion) domain of $0.7\!-\!1$~eV. The origin of both negative refraction bands can be understood from the power dissipation density by dielectric losses~\cite{Kittel}, which is proportional to $-\mbox{Im}(1/\varepsilon_y)$ shown by the green line in the figure, which exhibits two peaks representing the Drude and interband plasmons in their respective domains. Both of them come from the zeros of $\mbox{Re}\,\varepsilon_y$ as it changes its sign from $-$ to $+$ with energy increase to fulfil the Kramers-Kr\"{o}nig relation between  $\mbox{Re}\,\varepsilon_y$ and $\mbox{Im}\,\varepsilon_y$ as per the fundamental causality principle (opposite sign change sequence is representative of negative dispersion which gives the exciton absorption peak in $\mbox{Im}\,\varepsilon_y$).

In general~\cite{Ando2005}, due to the nanotube peculiar electronic band structure originating from the circumferential quantization of the longitudinal electron motion, SWCN real axial optical conductivities (given by the imaginary parts of the respective EM response functions along the SWCN symmetry axis) consist of the 1st, 2nd, 3rd, ... exciton resonances giving rise to a whole series of negative refraction bands like the one discussed above. SWCNs of different diameters and chiralities feature similar electronic band structure peculiarities, yet shifted in frequency relative to one another due to their diameter and chirality variation~\cite{Bond2012,Bondarev2015a}. This is why periodically aligned, homogeneous and weakly inhomogeneous TD films of SWCNs are excellent candidates for the development of multifunctional hyperbolic metasurfaces, to push the negative refraction and respective hyperbolic response (typically pertinent to the near-IR~\cite{Chen20}) into the optical spectral domain, with characteristics adjustable on demand by means of their composition (SWCN diameter, chirality, periodicity) and charge density (doping) variation.

Figure~\ref{fig3}~(a) presents the anisotropic LDOS of Eq.~(\ref{PDOS}) calculated with the EM field Green tensor of Eq.~(\ref{Greentensor}) as a function of the transition dipole relative orientation angles as sketched in Fig.~\ref{fig1}. The dipole emitter is positioned at $z_A\!=\!5,6\:\mbox{and}\:7$~nm with its excitation energy chosen to be $E\!=\!\hbar\omega_A\!\approx\!0.84$~eV right in the middle of the negative refraction band of the $\mbox{Re}\,\varepsilon_{y}$ function given by Eqs.~(\ref{EMtensor})--(\ref{plasmaFy}) and presented in Fig.~\ref{fig2}. It can be seen that the maximum of the anisotropic LDOS and the maximal spontaneous emission rate, accordingly, are to be expected in the plane perpendicular to the SWCN alignment along the direction perpendicular to the metasurface plane, in which case the transition dipole vector of the emitter is parallel to the SWCN alignment direction (see Fig.~\ref{fig1}). Getting the emitter closer to the film surface increases the LDOS in this particular direction, whereby the dipolar spontaneous emission rate increases unidirectionally to exceed the vacuum spontaneous emission rate by a factor $\sim\!10^6$ at $z_A\!=\!5$~nm as prescribed by Eq~(\ref{Gamma0}). Figure~\ref{fig3}~(b) shows the fixed-angle max LDOS [taken at $\vartheta\!=\!\varphi\!=\!\pi/2$ per Fig.~\ref{fig3}~(a)] as a function of energy $E\!=\!\hbar\omega$ and position $z_A$ of the dipole emitter. Here, by the comparison with Fig.~\ref{fig2}, it can be seen that the quantum negative refraction band of $\mbox{Re}\,\varepsilon_{y}$ results in the broad LDOS resonance (as broad as the negative refraction band is) with unidirectional spontaneous emission rate that exceeds the vacuum one by a factor $\sim\!10^8$ for $E\!=\!\hbar\omega_A\!\approx\!\hbar\omega_r$ (resonance condition) and $z_A\!=\!1$~nm. Clearly, this strong near-field effect is contributed predominantly by the evanescent part of the EM field Green tensor as can be seen from Eq.~(\ref{propevan}) with $l\!=\!l_A\!\sim\!0$.

\begin{figure}[t]
\hskip0.5cm\includegraphics[width=\linewidth]{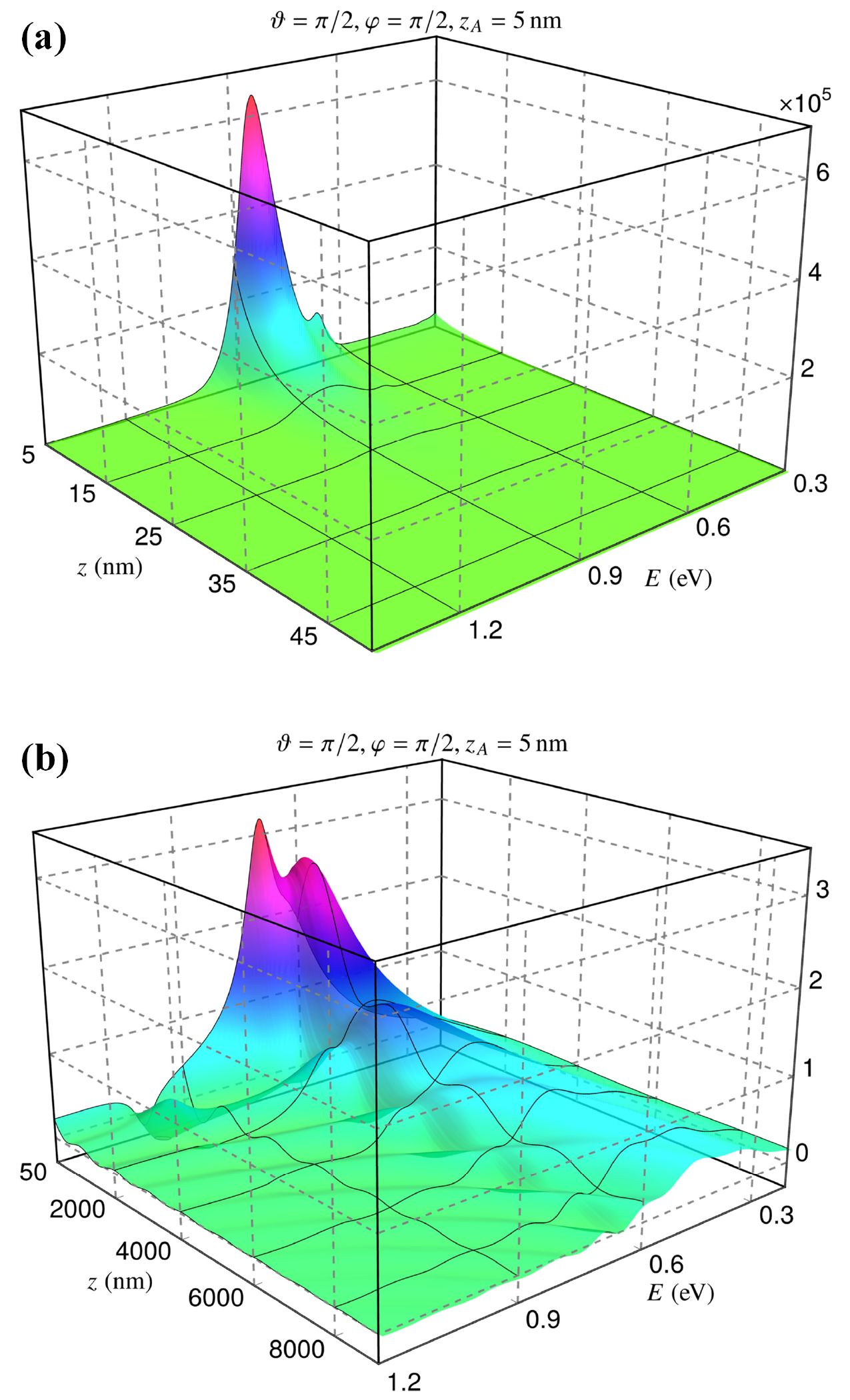}
\caption{Short-distance (a) and long-distance (b) fixed-angle max LDOS of Eq.~(\ref{xirar}) with dipole emitter position fixed at $z_A\!=\!5$~nm as functions of energy and observation point position. Both functions are calculated with EM field Green tensor of Eq.~(\ref{Greentensor}) in the form prescribed by Eq.~(\ref{propevan}) to separate out the evanescent and propagating wave contributions. In (a) both contributions are included; in (b) only the propagating contribution is included with its sign inverted.}
\label{fig4}
\end{figure}

Figure~\ref{fig4} shows the calculated short-distance (a) and long-distance (b) behaviors of the fixed-angle ($\vartheta\!=\!\varphi\!=\!\pi/2$) max LDOS function $\xi_y(z,z_A,\omega)$ of Eq.~(\ref{xirar}) for varying observation point (detector) position $z$ and varying energy $E\!=\!\hbar\omega$ with emitter position fixed at $z_A\!=\!5$~nm. As can be seen from Eqs.~(\ref{Ifin})--(\ref{xiws}), it is this function that totally controls the distance and time dependences of the anisotropic resonance fluorescence intensity of an excited dipole emitter in close proximity to the metasurface. Both graphs are calculated with the EM field Green tensor of Eq.~(\ref{Greentensor}) in the form prescribed by Eq.~(\ref{propevan}) in order to be able to see the role of the evanescent and propagating wave contributions separately as the observation point $z$ moves away from the metasurface structure. In (a) both terms are included in the calculation, while in (b) only the propagating term is included with its sign inverted. Clearly, as prescribed by Eq.~(\ref{propevan}), the evanescent (near-field wave) contribution dominates at short $z$ ($>\!z_A$) so much that the propagating wave contribution is totally hidden and cannot be seen in~(a). The short-distance behavior totally resembles that presented in Fig.~\ref{fig3}~(b) with all the same features discussed above. However, as the observation point moves away perpendicular to the metasurface plane, the dominant role played by the evanescent waves close to the surface is instead taken on by their propagating counterparts as given by Eq.~(\ref{propevan}) for large $z$ ($\gg\!z_A$) and shown in (b). Here, as $z$ increases, the function $\xi_y(z,z_A,\omega)$ can be inferred from the propagating part of Eq.~(\ref{propevan}) to oscillate at frequency $\sqrt{\epsilon_2}E/\hbar$, dropping down in amplitude as dictated by $\bar{R}_{yy}\big(\!\sqrt{1\!-\!y^2})|_{y\sim1}$ reflection coefficient contributing the most to the $z$-dependence of the integral and controlled by $\mbox{Re}\,\varepsilon_y$ and $\mbox{Re}\,\varepsilon_y$ shown in Fig.~\ref{fig2}. Similarly, too, does the anisotropic fluorescence intensity of the dipole emitter as can be seen from Eqs.~(\ref{Ifin})--(\ref{xiws}) with $I_0(z)\!\sim\!|\xi_y(z)|^2\!\gg\!|\xi_{x,z}|^2$, and as was first reported experimentally by Drexhage for (isotropic) atomic fluorescence intensity near a dielectric interface~\cite{Drexhage}. From the comparison of (a) and (b), it can also be seen that the fluorescence band broadens and shifts to the red with $z$ increasing.

The giant anisotropy and broad-band character of the spontaneous emission and fluorescence LDOS functions as well as the relative simplicity of the associated negative refraction band engineering by means such as SWCN chirality, diameter, composition variation~\cite{Bondarev2021} and simply by electron doping~\cite{Roberts2019}, open up routs for designing mechanically stable, ultrasensitive TD metasurfaces of aligned SWCNs on demand. As long as the frequency $\omega_A$ ($=\!\omega_{ul}-|\delta\omega_{ul}|$) of a dipole emitter fits into a negative refraction band of the SWCN metasurface, its spontaneous emission and fluorescence will be enhanced anisotropically as discussed above. At the very least, this remarkable property of ultrathin SWCN metasurfaces allows for very efficient design of single-atom optical sensors and solid-state single-photon sources in addition to those being developed with other 2D material platforms~\cite{Bond2022,Aharon2022}. Our theory can be tested on properly fabricated metasurfaces of small-diameter ($\sim\!1$~nm) periodically aligned SWCNs in experiments similar to those done with Eu$^{3+}$ ligands~\cite{Drexhage} or nanodimond NV-centers~\cite{SimeonBogdanov} as dipole emitters to be placed on a low-index superstrate of varied thickness deposited on the top of the SWCN film.

\section{Conclusions}

We use the medium-assisted QED approach to study the directionality effects in spontaneous emission and fluorescence of atoms (or molecules) in close proximity to finite-thickness ultrathin films of periodically aligned SWCNs presenting an important case of TD metasurfaces with nonlocal and highly anisotropic in-plane EM response. The EM response nonlocality is known to substantially modify nanomaterial EM properties~\cite{Shen2023,BiehsBond2023,Bond2022,Shah2022}. The nanomaterial system we study here restructures the spectral and spatial distribution of the EM eigenmodes pertaining to both virtual (near-field spontaneous emission) and real (far-field fluorescence) medium-assisted photon emission processes to enhance them unidirectionally. We show that, in addition to the strong Casimir effect anisotropy reported recently~\cite{Bondarev2023,Pablo2024}, atomic dipolar spontaneous emission and fluorescence can be enhanced by orders of magnitude in the plane perpendicular to the SWCN alignment along the direction perpendicular to the metasurface plane, in contrast to the commonly believed viewpoint of their uncontrollably random directionality. Thus, the alignment provides an extra measure for both near- and far-field photon emission control in addition to SWCN metasurface composition parameters. With further experimental development, the periodically aligned TD films of SWCNs can provide a multifunctional platform for an efficient hyperbolic metasurface design to cover both IR and visible spectral ranges for nanophotonics and quantum information science applications.

\acknowledgments

I.V.B. gratefully acknowledges support from the U.S. Army Research Office under award No.$\:$W911NF2310206. M.D.P. and S.F.I. were supported in part by the U.S. National Science Foundation grant No.$\;$DMR-1830874 awarded to I.V.B..

\appendix

\section{Derivation of Eq.~(\ref{IntensityK})}

We start by observing that Eq.~(\ref{Intensity}) can be written as
\begin{equation}
I(\textbf{r},t)=\!\!\!\sum_{\mu=x,y,z}\!\!\Big[\hat{E}^{(+)}_{\mu}(\mathbf{r})\hat{\sigma}^\dag|\psi(t)\rangle\Big]^\dag\hat{E}^{(+)}_{\mu}(\mathbf{r})\hat{\sigma}^{\dag}|\psi(t)\rangle,
\label{A1}
\end{equation}
whereby it is enough to only evaluate the rightmost factor. With Eqs.~(\ref{Ew})-(\ref{noise}) and Eq.~(\ref{wfunc}) this takes the form
\begin{eqnarray}
\hat{E}^{(+)}_{\mu}(\mathbf{r})\hat{\sigma}^{\dag}|\psi(t)\rangle=i\frac{2}{c^2}\!\int_{0}^{\infty}\!\!\!\!\!\!d\omega\!\!\int\!\!d\bm{\rho}\,\omega^2e^{-i\omega t}\hskip1.25cm
\label{A2}\\
\times\!\!\!\sum_{\lambda=x,y,z}\!\!\!C_{l\lambda}(\bm{\rho},\omega,t)\sqrt{\hbar d\,\mbox{Im}\,\varepsilon_{\lambda\lambda}(\bm{\rho},\omega)}\,G_{\mu\lambda}(\textbf{r},\bm{\rho},\omega)|\{0\}\rangle.\nonumber
\end{eqnarray}
Here, we used the following rules for the Pauli operators
\[
\hat{\sigma}^{\dag}|l\rangle=|u\rangle,\;\hat{\sigma}|u\rangle=|l\rangle,\;\hat{\sigma}^{\dag}|u\rangle=\hat{\sigma}|l\rangle=0
\]
as well as the commutation relations (\ref{fcommut}) to obtain
\[
\hat{f}_j(\bm{\rho}^\prime\!,\omega^\prime)\hat{f}^\dagger_\lambda(\bm{\rho},\omega)|\{0\}\rangle=|\{0\}\rangle\delta_{\lambda j}\delta(\bm{\rho}-\bm{\rho}^{\prime})\delta(\omega-\omega^{\prime}).
\]

Substituting into Eq.~(\ref{A2}) the result of the formal integration of Eq.~(\ref{popampl}),
\begin{eqnarray}
{C}_{l\lambda}(\bm{\rho},\omega,t)=\frac{2\omega^2}{\hbar c^2}\!\int_{0}^{t}\!\!\!d\tau\,e^{i(\omega-\omega_{A})\tau}C_{u}(\tau)\label{A3}\\
\times\!\!\!\sum_{\nu=x,y,z}\!\!\!\!\sqrt{\hbar d\,\mbox{Im}\,\varepsilon_{\lambda\lambda}(\bm{\rho},\omega)}\,d_{\nu}G_{\nu\lambda}^\ast(\textbf{r}_{A},\bm{\rho},\omega),\nonumber
\end{eqnarray}
and using the identity (\ref{mainidentity}), one further arrives at
\begin{eqnarray}
\hat{E}^{(+)}_{\mu}(\mathbf{r})\hat{\sigma}^{\dag}|\psi(t)\rangle=i\frac{4}{c^2}\!\int_{0}^{t}\!\!\!d\tau\,e^{-i\omega_{A}\tau}C_{u}(\tau)\hskip0.75cm\nonumber\\
\times\!\int_{0}^{\infty}\!\!\!\!\!\!d\omega\,\omega^2e^{-i\omega(t-\tau)}\!\!\!\!\sum_{\lambda=x,y,z}\!\!\!\mbox{Im}\,G_{\mu\lambda}(\textbf{r},\textbf{r}_A,\omega)d_\lambda|\{0\}\rangle,\nonumber
\end{eqnarray}
which being plugged into Eq.~(\ref{A1}) leads to Eq.~(\ref{IntensityK}).

\section{Derivation of Eq.~(\ref{Ifin})}

For $C_u(\tau)$ given by Eq.~(\ref{singleres}) the time-dependent factor in Eq.~(\ref{IntensityK}) takes the form
\begin{eqnarray}
\int_{0}^{t}\!\!\!d\tau e^{-i\omega(t-\tau)}C_u(\tau)e^{-i\omega_{A}\tau}\hskip1.5cm\label{B1}\\
=e^{-i\omega t}\Big[c_{+}\!\!\int_{0}^{t}\!\!\!d\tau\,e^{i(\omega-\omega_{-\!})\tau}\!+c_{-}\!\!\int_{0}^{t}\!\!\!d\tau\,e^{i(\omega-\omega_{+\!})\tau}\Big]\nonumber\\
=e^{-i\omega t}\Big[c_{+}\frac{e^{i(\omega-\omega_{-\!})t}-1}{i(\omega-\omega_{-\!})}+c_{-}\frac{e^{i(\omega-\omega_{+\!})t}-1}{i(\omega-\omega_{+\!})}\Big]\hskip0.15cm\nonumber
\end{eqnarray}\\
with $\omega_\pm\!=\omega_{A\!}-ic_\pm\delta\omega_r/(c_\pm\!-c_\mp)$. For our purposes here, this can further be treated in the long-time approximation using the well-known identity~\cite{DavydovQM}
\[
\lim_{t\rightarrow\infty}\frac{1-e^{-ixt}}{ix}=\pi\delta(x)-i{\cal{P}}\frac{1}{x}
\]
(${\cal{P}}$ denotes the principal value), whereby Eq.~(\ref{B1}) can be written as
\begin{eqnarray}
\int_{0}^{t}\!\!\!d\tau e^{-i\omega(t-\tau)}C_u(\tau)e^{-i\omega_{A}\tau}\Big|_{t\rightarrow\infty}\hskip0.5cm\label{B2}\\[0.25cm]
\approx\pi e^{-i\omega t}\big\{c_{+}\delta(\omega-\omega_{-})+c_{-}\delta(\omega-\omega_{+})\big\}\nonumber\\[0.25cm]
+\,ie^{-i\omega t}\,{\cal{P}}\Big(\frac{c_{+}}{\omega-\omega_{-}}+\frac{c_{-}}{\omega-\omega_{+}}\Big).\hskip0.75cm\nonumber
\end{eqnarray}
Plugging Eq.~(\ref{B2}) into Eq.~(\ref{IntensityK}) and observing that its final parenthesized term integrates over $\omega$ to zero due to the preceding alternating-sign exponential factor, one ultimately arrives at Eq.~(\ref{Ifin}).

\end{document}